\documentclass[preprint,11pt]{elsarticle}
\usepackage{graphicx,xcolor,amsmath,amssymb,amsthm,eufrak,physics,fullpage}
\usepackage[english]{babel}
\usepackage[hidelinks,colorlinks=true,linkcolor=blue,citecolor=blue,urlcolor=blue]{hyperref}
\biboptions{sort&compress}

\journal{{\tt arxiv.org}}
\newcommand{\eq}[2]{\begin{gather} #2  \label{#1} \end{gather}}
\newcommand{\nn}{\nonumber}
\newcommand{\Dt}{\mathcal{D}(t)}

\begin{document}

\begin{frontmatter}

\title{\bf On Gill's problem with anomalous diffusion} 

\author{\bf A. Barletta}

\address{\rm Department of Industrial Engineering, Alma Mater Studiorum Universit\`a di Bologna,\\
         Viale Risorgimento 2, Bologna 40136, Italy}

\begin{abstract}
The classical formulation of Gill's theorem establishing the linear stability of the  buoyant flow in a vertical porous channel subject to transverse diffusion and bounded by impermeable walls is here reconsidered from the perspective of anomalous mass diffusion. Anomalous diffusion is a physical effect implying a more general formulation of the governing equations of mass transfer either in a fluid or in a fluid-saturated porous medium. Anomalous diffusion is mathematically modelled via a diffusion equation with a time-dependent mass diffusivity. In this study, the stationary and parallel buoyant flow in a vertical porous layer saturated by a fluid is investigated. The implications of the anomalous diffusion model for the proof that such a flow is always linearly stable are analysed. 
\end{abstract}

\end{frontmatter}

\section{Introduction}
Diffusion is among the oldest and most successful paradigms of transport theory. Einstein's analysis of Brownian motion \citep{Einstein1905} established that, for a particle undergoing many small, independent and identically distributed displacements, the mean square displacement grows linearly in time, $\langle x^2(t)\rangle \sim \mathcal{D}t$, and that the solute concentration $C$ obeys the diffusion partial differential equation,
\begin{equation}
  \pdv{C}{t} = \mathcal{D}\,\laplacian{C} .
  \label{eq:fick}
\end{equation}
Here, $x^2$ is the square displacement of the particle, $t$ is time and $\mathcal{D}$ is the diffusion coefficient.

In many natural and engineered systems, however, this picture fails. Solutes in aquifers, tracers in fractured rock, charge carriers in amorphous semiconductors and particles in crowded biological media may display transport that is not described by \eqref{eq:fick} thus indicating an anomaly in the diffusion process \citep{BouchaudGeorges1990,Sokolov2002,Klages2008}.
 
Diffusion is called anomalous when the mean square displacement departs from linear growth. A typical representation of experimental data is given by
\begin{equation}
  \langle x^2(t)\rangle \sim \mathcal{D}_\alpha\, t^{\alpha}, \qquad \alpha\neq 1,
  \label{eq:msd}
\end{equation}
with $0<\alpha<1$ for subdiffusion and $\alpha>1$ for superdiffusion \citep{MetzlerKlafter2000}. We mention that, by setting $\alpha = 1$ (with $\mathcal{D}_\alpha = \mathcal{D}$), \eqref{eq:msd} agrees with \eqref{eq:fick} thus including normal diffusion as a special case. An early macroscopic example of anomalous diffusion is Richardson's observation that the relative dispersion of particles in turbulent flow grows faster than for normal diffusion \citep{Richardson1926}. The models of anomalous diffusion show how to include these ideas in the study of convection and flow instability in porous media.                                                                                                                                                                                                                                                                                                                                                                                                                                                                                                                                                                                                                                                                                                                                                                                                                                                                                                                                                                                                                                                                       
 
Two microscopic mechanisms are usually invoked for anomalous diffusion. In the first, particles are trapped leading to subdiffusion. This phenomenon was formalised in the continuous time random walk framework of \citet{MontrollWeiss1965}, later used by Scher and Montroll to explain dispersive transport in disordered solids \citep{ScherMontroll1975}. In porous media, trapping occurs in stagnant zones, dead-end pores and low-permeability inclusions. In the second mechanism, jump lengths have a heavy-tailed distribution giving L\'evy flights and superdiffusion. In aquifers, this phenomenology corresponds to rapid transport along connected high-conductivity channels \citep{Benson2000}. 
A comprehensive review of both these mechanisms is given in \citet{BouchaudGeorges1990} and in \citet{MetzlerKlafter2000}. A model for anomalous diffusion is based on a time-dependent mass diffusivity, $\Dt$, which may have the form of a power-law model based on \eqref{eq:msd}, widely employed in the correlation of experimental data, or it can address more general forms of time dependence \cite{henry2010introduction}.

An anomalous diffusion equation with a time-dependent diffusivity, $\Dt$, has been employed recently to establish its impact on the onset of Rayleigh-B\'enard instability in porous media \cite{barletta2023rayleigh, straughan2024asymptotic, BARLETTA2025109410, BarReCeBr2025, barletta2026prats}. All these studies pointed out the stabilising effect of superdiffusion and the destabilising effect of subdiffusion. The interplay of anomalous diffusion and convective instability in porous media was previously explored by other authors \cite{karani2017onset, klimenko2017effect} who based their studies on fractional differential calculus models.

In his classical paper, Gill proved that the parallel and stationary buoyant flow in a vertical porous layer with impermeable boundaries cannot undergo any transition to linear instability \cite{gill1969proof}. Physically, this result implies that the buoyant flow cannot break up into multicellular flow patterns, as it happens for a fluid layer in a vertical cavity \cite{vest1969stability}. Gill's paper is about heat diffusion instead of mass diffusion, but this aspect has has no bearing on the results of his analysis, because the mathematical models employed for the two types of diffusion are formally identical.

Later on, Gill's theorem was proved to hold also when nonlinear perturbations are devised \cite{Stra88} or when local non-equilibrium effects in the diffusion process occur \cite{Rees11, scott2013nonlinear, straughan2015micro}. On the other hand, other authors have recently shown that flow instability may emerge when either the boundaries are permeable \cite{Barletta2015} or when the porous medium is heterogeneous \cite{shankar2022gill}. Gill's analysis \cite{gill1969proof} and all the later investigations on this topic were carried out according to the model of normal diffusion.

The aim of this paper is to employ the methodology developed in Refs.~\cite{barletta2023rayleigh, barletta2026prats} to investigate the influence of the anomalous mass diffusion on the stability of the buoyancy-induced flow in a vertical porous layer with impermeable boundaries. In particular, we are interested in checking whether and how Gill's no-go theorem \cite{gill1969proof} for the instability of the parallel vertical flow in the porous layer still holds when the diffusion process driving the buoyancy force is anomalous. Section~\ref{sec2} presents the mathematical model and governing equations for buoyant Darcy's flow with anomalous mass diffusion in a vertical porous layer, together with their dimensionless scaling. Section~\ref{sec3} derives the equilibrium solution of the governing equations corresponding to the parallel buoyant flow in the porous medium. Section~\ref{sec4} carries out the linear stability analysis of the equilibrium flow and proves the validity of Gill's no-go theorem for the instability. The conclusions are summarised in Section~\ref{sec5}.

\section{Mathematical model}\label{sec2}
\begin{figure}[t]
\centering
\includegraphics[width=0.43\textwidth]{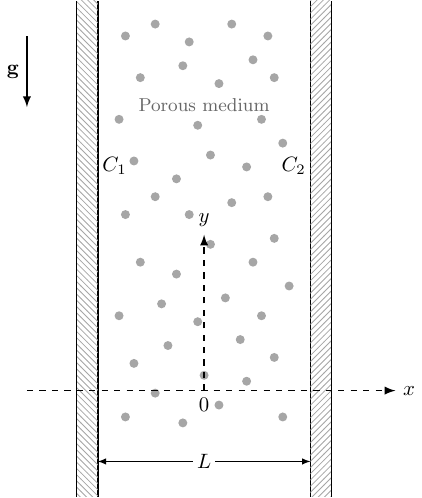}
\caption{\label{fig1}A two-dimensional sketch of the porous layer and of its boundaries in the $xy$ plane}
\end{figure}

A vertical porous layer saturated by a fluid binary mixture, such that $C$ is the concentration of the solute, while $\vb{u}$ is the velocity field of the fluid and $P$ is the dynamic pressure, namely the local difference between the pressure and the hydrostatic pressure. Along the $(x,y,z)$ axes, the Cartesian components of $\vb{u}$ are denoted by $(u,v,w)$. As sketched in Fig.~\ref{fig1}, the vertical boundary planes, $x=-L/2$ and $x=L/2$ feature constant concentrations $C_1$ and $C_2$, respectively. Such vertical boundaries are modelled as impermeable. 

\subsection{Governing equations}
The Oberbeck-Boussinesq approximation is adopted to model the buoyancy-induced flow \citep{straughan2008stability, NiBe17, barletta2022boussinesq}. Furthermore, the momentum transfer across the porous medium is described by Darcy's law, with the reference fluid density $\rho_0$ evaluated for $C = C_0$, where $C_0$ is the mean value $\qty(C_1 + C_2)/2$.
\eq{1}{
\div{\vb{u}} = 0,\nn\\
\frac{\mu}{K} \vb{u} = - \grad{P} - \rho_0\, \beta\, \qty(C - C_0)\, \vb{g},\nn\\
\varphi\pdv{C}{t} + \vb{u} \vdot \grad{C} = \varphi \,\Dt \laplacian{C}.
}
A time-dependent mass diffusivity relative to the fluid binary mixture, $\Dt$, is employed to model anomalous diffusion in the porous layer \citep{henry2010introduction, liang2021non, BARLETTA2025109410, BarReCeBr2025, girelli2025dynamical, barletta2026prats}. 
In \eqref{1}, $K$, $\varphi$, $\mu$, $\beta$ and $\vb{g}$ denote the permeability, porosity, dynamic viscosity, concentration expansion coefficient and gravity acceleration $\vb{g} = - g\, \vb{e}_y$. Here, $g$ is the modulus of $\vb{g}$ and $\vb{e}_y$ is the unit vector of the vertical $y$ axis.

We define the reference value, $\mathcal{D}_0$, of the time-dependent mass diffusivity, $\Dt$, so that $\mathcal{D}_0$ yields the limit of $\Dt$ at large times, for normal diffusion. On the other hand, for subdiffusion or superdiffusion, $\mathcal{D}_0$ can be defined so that $L^2/\mathcal{D}_0$ is a characteristic reference time for the anomalous diffusion process.

\subsection{Dimensionless scaling}

By scaling the coordinates, time and fields that define the physical system under consideration,
\eq{2}{
\frac{\qty(x,y,z)}{L} \to \qty(x ,  y ,  z) \qc 
\frac{t}{L^2/ \mathcal{D}_0} \to t \qc
 \frac{\vb{u}}{\varphi \mathcal{D}_0/L} \to \vb{u} , \nn\\
\frac{P}{\varphi \mu \mathcal{D}_0/K} \to P \qc \frac{C - C_0}{C_2 - C_1} \to C \qc \frac{\Dt}{\mathcal{D}_0} \to \Dt,
}
we obtain a dimensionless reformulation of \eqref{1},
\eq{3}{
\div{\vb{u}} = 0,\nn\\
\vb{u} = - \grad{P}  + Ra\, C\,  \vb{e}_y ,\nn\\
\pdv{C}{t} + \vb{u} \vdot \grad{C} = \Dt \laplacian{C}.
}
In \eqref{3}, $Ra$ is the Rayleigh number defined as
\eq{4}{
Ra = \frac{\rho_0 \beta g \qty(C_2 - C_1) K L}{\varphi \mu \mathcal{D}_0} . 
}
The dimensionless boundary conditions can be expressed as
\eq{5}{
x=\pm\, 1/2 : \qquad u = 0 \qc C = \pm\, \frac{1}{2}.
}

\section{The equilibrium solution}\label{sec3}
A stationary solution of \eqref{3} and \eqref{5} is given by
\eq{6}{
u_{eq} = 0 \qc v_{eq} = Ra\, x \qc w_{eq} = 0 \qc \grad{P_{eq}} = 0 \qc C_{eq} = x .
}
Here, subscript $eq$ indicates the equilibrium state. Such an equilibrium state describes a purely vertical motion of the saturating fluid, which is induced by the buoyancy force and which is endowed with a zero net flow-rate. We note that this equilibrium solution is in no way influenced by the possibly anomalous diffusion process and, as such, it is in perfect agreement with the basic flow state considered in the classical paper by \citet{gill1969proof}.

\section{Linear stability}\label{sec4}
We perturb the equilibrium state with disturbances having amplitude $\order{\epsilon}$ with $|\epsilon| \ll 1$,
\eq{7}{
\vb{u} = \qty(0, Ra\, x, 0) + \epsilon\, \tilde{\vb{u}} \qc \grad{P} = \epsilon\, \grad{\tilde{P}} \qc C = x + \epsilon\, \tilde{C},
}
where tilde denotes the perturbation field. By keeping only terms up to  $\order{\epsilon}$, the substitution of \eqref{7} into \eqref{3} and \eqref{5} yields
\eq{8}{
\div{\tilde{\vb{u}}} = 0,\nn\\
\tilde{\vb{u}} = - \grad{\tilde{P}}  + Ra\, \tilde{C}\,  \vb{e}_y ,\nn\\
\pdv{\tilde{C}}{t} + Ra\,x\, \pdv{\tilde{C}}{y} + \tilde{u} = \Dt \laplacian{\tilde{C}}, \nn\\
x=\pm\, 1/2 : \qquad \tilde{u} = 0 \qc \tilde{C} = 0.
}
Then, one may adopt a pressure-concentration restatement of \eqref{8} where
\eq{9}{
\laplacian{\tilde{P}}  = Ra\, \pdv{\tilde{C}}{y} ,\nn\\
\pdv{\tilde{C}}{t} + Ra\,x\, \pdv{\tilde{C}}{y} - \pdv{\tilde{P}}{x} = \Dt \laplacian{\tilde{C}}, \nn\\
x=\pm\, 1/2 : \qquad \pdv{\tilde{P}}{x} = 0 \qc \tilde{C} = 0.
}
Let us introduce normal mode expressions of $(\tilde{P},\tilde{C})$,
\eq{10}{
\qty{\mqty{\tilde{P}(x,y,z,t)\\ \tilde{C}(x,y,z,t)}} = \qty{\mqty{f(x,t)\\ h(x,t)}}\, e^{i(k_y y + k_z z)}  .
}
On account of \eqref{10}, \eqref{9} yields the differential problem
\eq{11}{
\pdv[2]{f}{x} - k^2 f  - ik_y Ra\, h = 0 ,\nn\\
\pdv{h}{t} + ik_y Ra x\, h - \pdv{f}{x} = \Dt \qty(\pdv[2]{h}{x} - k^2 h), \nn\\
x=\pm\, 1/2 : \qquad \pdv{f}{x} = 0 \qc h = 0,
}
with the wavenumber $k$ defined as $k = (k_y^2 + k_z^2)^{1/2}$. One may define a rescaled Rayleigh number,
\eq{12}{
R = Ra\, \frac{k_y}{k},
}
so that \eqref{11} yields
\eq{13}{
\pdv[2]{f}{x} - k^2 f - i k R\, h = 0 ,\nn\\
\pdv{h}{t} + ikRx\, h - \pdv{f}{x} = \Dt \qty(\pdv[2]{h}{x} - k^2 h), \nn\\
x=\pm\, 1/2 : \qquad \pdv{f}{x} = 0 \qc h = 0.
}
The two partial differential equations \eqref{13} can be combined into a single higher-order partial differential equation in $h$,
\eq{14}{
\Dt\qty(\pdv[4]{h}{x} - 2k^2\,\pdv[2]{h}{x} + k^4 h )
+ k^2\,\pdv{h}{t} - ikR \pdv{}{x} \qty( x\,\pdv{h}{x}) - \frac{\partial^3 h}{\partial x^2 \partial t} + ik^3Rx\,h = 0 ,\nn\\
x=\pm\, 1/2 : \qquad h = 0 \qc \pdv[2]{h}{x} = 0.
}
Let us multiply the differential equation \eqref{14} by $\bar{h}(x,t)$, the complex conjugate of $h(x,t)$, integrate over $x \in [-1/2,1/2]$ and use the boundary conditions \eqref{14} to develop integration by parts, 
\eq{15}{
\Dt \qty(\int\limits_{-1/2}^{1/2} \qty|\pdv[2]{h}{x}|^2\, \dd x + 2k^2 \int\limits_{-1/2}^{1/2} \qty|\pdv{h}{x}|^2\, \dd x + k^4 \int\limits_{-1/2}^{1/2} |h|^2\, \dd x )
+ k^2 \int\limits_{-1/2}^{1/2} \bar{h}\, \pdv{h}{t}\, \dd x \nn\\
+ ikR \int\limits_{-1/2}^{1/2} x \qty|\pdv{h}{x}|^2 \, \dd x + \int\limits_{-1/2}^{1/2} \pdv{\bar{h}}{x} \pdv{h}{x}{t}\, \dd x + ik^3R\int\limits_{-1/2}^{1/2} x |h|^2\, \dd x = 0 .
}
By evaluating the real part of \eqref{15}, one obtains
\eq{16}{
\dv{}{t} \qty( \,\int\limits_{-1/2}^{1/2} \qty|\pdv{h}{x}|^2\, \dd x + k^2 \int\limits_{-1/2}^{1/2} \qty|h|^2\, \dd x ) \nn\\
= - 2\Dt \qty(\,\int\limits_{-1/2}^{1/2} \qty|\pdv[2]{h}{x}|^2\, \dd x + 2k^2 \int\limits_{-1/2}^{1/2} \qty|\pdv{h}{x}|^2\, \dd x + k^4 \int\limits_{-1/2}^{1/2} |h|^2\, \dd x )  .
}
Whatever is the type of mass diffusion, the time-dependent diffusivity $\Dt$ has to be positive. As a consequence, \eqref{16} allows one to conclude that the non-negative function
\eq{17}{
F(t) = \int\limits_{-1/2}^{1/2} \qty|\pdv{h}{x}|^2\, \dd x +
k^2 \int\limits_{-1/2}^{1/2} \qty|h|^2\, \dd x
}
decreases in time. Hence, $|h(x,t)|$ cannot grow unboundedly at large times and the same feature holds also for the perturbation $\tilde{C}(x,y,z,t)$ of the solute concentration, on account of \eqref{10}. Finally, the first equation \eqref{9} leads to the deduction that also the pressure perturbation $\tilde{P}(x,y,z,t)$ cannot grow unboundedly at large times. 

The conclusion is that no linear instability can be developed for either subdiffusion, normal diffusion or superdiffusion.

The distinction between different regimes of mass diffusion is summarised as
\eq{18}{
\lim_{t \to +\infty} \Dt =
\begin{cases}
0 \qc &\text{for subdiffusion},\\
1 \qc &\text{for normal diffusion},\\
+\infty \qc &\text{for superdiffusion}.
\end{cases}
}
Therefore, \eqref{16} and \eqref{18} depict a scenario where different regimes of mass diffusion imply different rates of decay for function $F(t)$ defined by \eqref{17}. The largest decay rates are associated with superdiffusion, while the smallest occur for subdiffusion. The case of normal diffusion is a threshold regime where the decay rate is eventually exponential, as one can infer from \eqref{13}. In fact, at large times and with normal diffusion, the solution of \eqref{13} can be undertaken by employing separation of variables where the time dependence is embodied by a factor $e^{\lambda t}$ with $\Re(\lambda)<0$ \citep{gill1969proof}. In particular, if we write
\eq{19}{
h(x,t) = e^{\lambda t} H(x),
}
then \eqref{16} and \eqref{18} yield
\eq{20}{
\Re(\lambda) = - \frac{\displaystyle \int\limits_{-1/2}^{1/2} \qty|\dv[2]{H}{x}|^2\, \dd x + 2k^2 \int\limits_{-1/2}^{1/2} \qty|\dv{H}{x}|^2\, \dd x + k^4 \int\limits_{-1/2}^{1/2} |H|^2\, \dd x}{\displaystyle \int\limits_{-1/2}^{1/2} \qty|\dv{H}{x}|^2\, \dd x + k^2 \int\limits_{-1/2}^{1/2} \qty|H|^2\, \dd x} .
}
The asymptotic stability response \citep{khalil2002nonlinear}, implied by \eqref{19} and \eqref{20}, entails that the initial perturbation, with normal diffusion, will decay and eventually tend to zero for $t \to +\infty$. On the basis of \eqref{18}, the ultimate extinction of the initial perturbation is predicted {\em a fortiori} for superdiffusion. On the other hand, subdiffusion suggests the possibility of a scenario where $\dd F(t)/\dd t \to 0$ when $t \to +\infty$, where $F(t)$ is given by \eqref{17}. In practice, this means that the perturbation $h(x,t)$ will not necessarily tend to zero at large times. This scenario means stability, according to Lyapounov definition (see, for instance, Chapter 4 of the book by \citet{khalil2002nonlinear}), even if asymptotic stability does not necessarily occur.

\section{Conclusions}\label{sec5}
The effects of anomalous diffusion on the stability of the equilibrium parallel flow induced by the buoyancy force in a vertical porous layer saturated by a fluid have been analysed. The classical no-go theorem proved by Gill~\cite{gill1969proof}, showing that the instability of the equilibrium flow with normal diffusion is impossible, has been extended to include processes of subdiffusion and superdiffusion. As in the original paper by Gill~\cite{gill1969proof}, impermeability boundary conditions have been assumed, together with uniform Dirichlet boundary conditions for the advection/diffusion equation (heat diffusion in the case studied by Gill and mass diffusion in the present study). The model of anomalous mass diffusion employed in this paper is based on a time-dependent diffusivity coefficient, whose limiting behaviour at large times characterises the subdiffusion/superdiffusion regimes.

It has been shown analytically that the linear instability is impossible both for normal diffusion and for anomalous diffusion. This result marks a difference with respect to previous analyses of the onset of Rayleigh-B\'enard instability in a horizontal porous layer~\cite{barletta2023rayleigh, straughan2024asymptotic, BARLETTA2025109410, BarReCeBr2025, barletta2026prats}. In fact, the Rayleigh-B\'enard equilibrium state in a horizontal porous layer turns out to be always stable for superdiffusion and unstable for subdiffusion, with every positive Rayleigh number. In a horizontal layer, a positive Rayleigh number means a potentially unstable density stratification in the equilibrium state. 

In the case of Gill's problem considered in this paper, the anomalous (subdiffusive or superdiffusive) or normal nature of the mass diffusion process makes no difference relative to the impossibility of small-scale perturbations to grow in time, at large times. However, a difference may emerge as asymptotic stability shows up for normal diffusion and superdiffusion. On the other hand, stability (in the Lyapounov sense), not necessarily asymptotic, has been proved for subdiffusion.  

\section*{Acknowledgements}
This work was financially supported by Alma Mater Studiorum Universit\`a di Bologna,
grant number RFO-2025.


\begin{thebibliography}{31}
\expandafter\ifx\csname natexlab\endcsname\relax\def\natexlab#1{#1}\fi
\providecommand{\url}[1]{\texttt{#1}}
\providecommand{\href}[2]{#2}
\providecommand{\path}[1]{#1}
\providecommand{\DOIprefix}{doi:}
\providecommand{\ArXivprefix}{arXiv:}
\providecommand{\URLprefix}{URL: }
\providecommand{\Pubmedprefix}{pmid:}
\providecommand{\doi}[1]{\href{http://dx.doi.org/#1}{\path{#1}}}
\providecommand{\Pubmed}[1]{\href{pmid:#1}{\path{#1}}}
\providecommand{\bibinfo}[2]{#2}
\ifx\xfnm\relax \def\xfnm[#1]{\unskip,\space#1}\fi
\bibitem[{Einstein(1905)}]{Einstein1905}
\bibinfo{author}{A.~Einstein},
\newblock \bibinfo{title}{{\"U}ber die von der molekularkinetischen {T}heorie
  der {W}{\"a}rme geforderte {B}ewegung von in ruhenden {F}l{\"u}ssigkeiten
  suspendierten {T}eilchen},
\newblock \bibinfo{journal}{Annalen der Physik} \bibinfo{volume}{322}
  (\bibinfo{year}{1905}) \bibinfo{pages}{549--560}.
\bibitem[{Bouchaud and Georges(1990)}]{BouchaudGeorges1990}
\bibinfo{author}{J.-P. Bouchaud}, \bibinfo{author}{A.~Georges},
\newblock \bibinfo{title}{Anomalous diffusion in disordered media: Statistical
  mechanisms, models and physical applications},
\newblock \bibinfo{journal}{Physics Reports} \bibinfo{volume}{195}
  (\bibinfo{year}{1990}) \bibinfo{pages}{127--293}.
\bibitem[{Sokolov et~al.(2002)Sokolov, Klafter, and Blumen}]{Sokolov2002}
\bibinfo{author}{I.~M. Sokolov}, \bibinfo{author}{J.~Klafter},
  \bibinfo{author}{A.~Blumen},
\newblock \bibinfo{title}{Fractional kinetics},
\newblock \bibinfo{journal}{Physics Today} \bibinfo{volume}{55}
  (\bibinfo{year}{2002}) \bibinfo{pages}{48--54}.
\bibitem[{Klages et~al.(2008)Klages, Radons, and Sokolov}]{Klages2008}
\bibinfo{editor}{R.~Klages}, \bibinfo{editor}{G.~Radons},
  \bibinfo{editor}{I.~M. Sokolov} (Eds.), \bibinfo{title}{Anomalous Transport:
  Foundations and Applications}, \bibinfo{publisher}{Wiley-VCH},
  \bibinfo{address}{Weinheim}, \bibinfo{year}{2008}.
\bibitem[{Metzler and Klafter(2000)}]{MetzlerKlafter2000}
\bibinfo{author}{R.~Metzler}, \bibinfo{author}{J.~Klafter},
\newblock \bibinfo{title}{The random walk's guide to anomalous diffusion: a
  fractional dynamics approach},
\newblock \bibinfo{journal}{Physics Reports} \bibinfo{volume}{339}
  (\bibinfo{year}{2000}) \bibinfo{pages}{1--77}.
\bibitem[{Richardson(1926)}]{Richardson1926}
\bibinfo{author}{L.~F. Richardson},
\newblock \bibinfo{title}{Atmospheric diffusion shown on a distance-neighbour
  graph},
\newblock \bibinfo{journal}{Proceedings of the Royal Society of London. Series
  A} \bibinfo{volume}{110} (\bibinfo{year}{1926}) \bibinfo{pages}{709--737}.
\bibitem[{Montroll and Weiss(1965)}]{MontrollWeiss1965}
\bibinfo{author}{E.~W. Montroll}, \bibinfo{author}{G.~H. Weiss},
\newblock \bibinfo{title}{Random walks on lattices. {II}},
\newblock \bibinfo{journal}{Journal of Mathematical Physics}
  \bibinfo{volume}{6} (\bibinfo{year}{1965}) \bibinfo{pages}{167--181}.
\bibitem[{Scher and Montroll(1975)}]{ScherMontroll1975}
\bibinfo{author}{H.~Scher}, \bibinfo{author}{E.~W. Montroll},
\newblock \bibinfo{title}{Anomalous transit-time dispersion in amorphous
  solids},
\newblock \bibinfo{journal}{Physical Review B} \bibinfo{volume}{12}
  (\bibinfo{year}{1975}) \bibinfo{pages}{2455--2477}.
\bibitem[{Benson et~al.(2000)Benson, Wheatcraft, and Meerschaert}]{Benson2000}
\bibinfo{author}{D.~A. Benson}, \bibinfo{author}{S.~W. Wheatcraft},
  \bibinfo{author}{M.~M. Meerschaert},
\newblock \bibinfo{title}{Application of a fractional advection-dispersion
  equation},
\newblock \bibinfo{journal}{Water Resources Research} \bibinfo{volume}{36}
  (\bibinfo{year}{2000}) \bibinfo{pages}{1403--1412}.
\bibitem[{Henry et~al.(2010)Henry, Langlands, and
  Straka}]{henry2010introduction}
\bibinfo{author}{B.~I. Henry}, \bibinfo{author}{T.~A.~M. Langlands},
  \bibinfo{author}{P.~Straka},
\newblock \bibinfo{title}{An introduction to fractional diffusion},
\newblock in: \bibinfo{editor}{R.~L. Dewar}, \bibinfo{editor}{F.~Detering}
  (Eds.), \bibinfo{booktitle}{Complex Physical, Biophysical and Econophysical
  Systems}, \bibinfo{publisher}{World Scientific},
  \bibinfo{address}{Singapore}, \bibinfo{year}{2010}, pp.
  \bibinfo{pages}{37--89}.
\bibitem[{Barletta(2023)}]{barletta2023rayleigh}
\bibinfo{author}{A.~Barletta},
\newblock \bibinfo{title}{Rayleigh-{B\'e}nard instability in a horizontal
  porous layer with anomalous diffusion},
\newblock \bibinfo{journal}{Physics of Fluids} \bibinfo{volume}{35}
  (\bibinfo{year}{2023}) \bibinfo{pages}{104114}.
\bibitem[{Straughan and Barletta(2024)}]{straughan2024asymptotic}
\bibinfo{author}{B.~Straughan}, \bibinfo{author}{A.~Barletta},
\newblock \bibinfo{title}{Asymptotic behaviour for convection with anomalous
  diffusion},
\newblock \bibinfo{journal}{Continuum Mechanics and Thermodynamics}
  \bibinfo{volume}{36} (\bibinfo{year}{2024}) \bibinfo{pages}{737--743}.
\bibitem[{Barletta et~al.(2025{\natexlab{a}})Barletta, Brandão, Capone, and
  {De Luca}}]{BARLETTA2025109410}
\bibinfo{author}{A.~Barletta}, \bibinfo{author}{P.~V. Brandão},
  \bibinfo{author}{F.~Capone}, \bibinfo{author}{R.~{De Luca}},
\newblock \bibinfo{title}{Stabilizing effect of generic anomalous diffusion
  independent of the {R}ayleigh number},
\newblock \bibinfo{journal}{International Communications in Heat and Mass
  Transfer} \bibinfo{volume}{168} (\bibinfo{year}{2025}{\natexlab{a}})
  \bibinfo{pages}{109410}.
\bibitem[{Barletta et~al.(2025{\natexlab{b}})Barletta, Rees, Celli, and
  {Brand\~ao}}]{BarReCeBr2025}
\bibinfo{author}{A.~Barletta}, \bibinfo{author}{D.~A.~S. Rees},
  \bibinfo{author}{M.~Celli}, \bibinfo{author}{P.~V. {Brand\~ao}},
\newblock \bibinfo{title}{Open boundaries, anomalous diffusion and the
  {D}arcy‐{B\'e}nard instability},
\newblock \bibinfo{journal}{Transport in Porous Media} \bibinfo{volume}{152}
  (\bibinfo{year}{2025}{\natexlab{b}}) \bibinfo{pages}{27}.
\bibitem[{Barletta(2026)}]{barletta2026prats}
\bibinfo{author}{A.~Barletta},
\newblock \bibinfo{title}{On {P}rats’ problem with anomalous diffusion},
\newblock \bibinfo{journal}{Journal of Fluid Mechanics} \bibinfo{volume}{1032}
  (\bibinfo{year}{2026}) \bibinfo{pages}{A18}.
\bibitem[{Karani et~al.(2017)Karani, Rashtbehesht, Huber, and
  Magin}]{karani2017onset}
\bibinfo{author}{H.~Karani}, \bibinfo{author}{M.~Rashtbehesht},
  \bibinfo{author}{C.~Huber}, \bibinfo{author}{R.~L. Magin},
\newblock \bibinfo{title}{Onset of fractional-order thermal convection in
  porous media},
\newblock \bibinfo{journal}{Physical Review E} \bibinfo{volume}{96}
  (\bibinfo{year}{2017}) \bibinfo{pages}{063105}.
\bibitem[{Klimenko and Maryshev(2017)}]{klimenko2017effect}
\bibinfo{author}{L.~S. Klimenko}, \bibinfo{author}{B.~S. Maryshev},
\newblock \bibinfo{title}{Effect of solute immobilization on the stability
  problem within the fractional model in the solute analog of the
  {H}orton-{R}ogers-{L}apwood problem},
\newblock \bibinfo{journal}{The European Physical Journal E}
  \bibinfo{volume}{40} (\bibinfo{year}{2017}) \bibinfo{pages}{1--7}.
\bibitem[{Gill(1969)}]{gill1969proof}
\bibinfo{author}{A.~E. Gill},
\newblock \bibinfo{title}{A proof that convection in a porous vertical slab is
  stable},
\newblock \bibinfo{journal}{Journal of Fluid Mechanics} \bibinfo{volume}{35}
  (\bibinfo{year}{1969}) \bibinfo{pages}{545--547}.
\bibitem[{Vest and Arpaci(1969)}]{vest1969stability}
\bibinfo{author}{C.~M. Vest}, \bibinfo{author}{V.~S. Arpaci},
\newblock \bibinfo{title}{Stability of natural convection in a vertical slot},
\newblock \bibinfo{journal}{Journal of Fluid Mechanics} \bibinfo{volume}{36}
  (\bibinfo{year}{1969}) \bibinfo{pages}{1--15}.
\bibitem[{Straughan(1988)}]{Stra88}
\bibinfo{author}{B.~Straughan},
\newblock \bibinfo{title}{{A nonlinear analysis of convection in a porous
  vertical slab}},
\newblock \bibinfo{journal}{Geophysical {\&} Astrophysical Fluid Dynamics}
  \bibinfo{volume}{42} (\bibinfo{year}{1988}) \bibinfo{pages}{269--275}.
\bibitem[{Rees(2011)}]{Rees11}
\bibinfo{author}{D.~A.~S. Rees},
\newblock \bibinfo{title}{The effect of local thermal nonequilibrium on the
  stability of convection in a vertical porous channel},
\newblock \bibinfo{journal}{Transport in Porous Media} \bibinfo{volume}{87}
  (\bibinfo{year}{2011}) \bibinfo{pages}{459--464}.
\bibitem[{Scott and Straughan(2013)}]{scott2013nonlinear}
\bibinfo{author}{N.~L. Scott}, \bibinfo{author}{B.~Straughan},
\newblock \bibinfo{title}{A nonlinear stability analysis of convection in a
  porous vertical channel including local thermal nonequilibrium},
\newblock \bibinfo{journal}{Journal of Mathematical Fluid Mechanics}
  \bibinfo{volume}{15} (\bibinfo{year}{2013}) \bibinfo{pages}{171--178}.
\bibitem[{Straughan(2015)}]{straughan2015micro}
\bibinfo{author}{B.~Straughan}, \bibinfo{title}{Convection with Local Thermal
  Non-Equilibrium and Microfluidic Effects}, \bibinfo{publisher}{Springer},
  \bibinfo{address}{New York, NY}, \bibinfo{year}{2015}.
\bibitem[{Barletta(2015)}]{Barletta2015}
\bibinfo{author}{A.~Barletta},
\newblock \bibinfo{title}{A proof that convection in a porous vertical slab may
  be unstable},
\newblock \bibinfo{journal}{Journal of Fluid Mechanics} \bibinfo{volume}{770}
  (\bibinfo{year}{2015}) \bibinfo{pages}{273--288}.
\bibitem[{Shankar and Shivakumara(2022)}]{shankar2022gill}
\bibinfo{author}{B.~M. Shankar}, \bibinfo{author}{I.~S. Shivakumara},
\newblock \bibinfo{title}{Gill's stability problem may be unstable with
  horizontal heterogeneity in permeability},
\newblock \bibinfo{journal}{Journal of Fluid Mechanics} \bibinfo{volume}{943}
  (\bibinfo{year}{2022}) \bibinfo{pages}{A20}.
\bibitem[{Straughan(2008)}]{straughan2008stability}
\bibinfo{author}{B.~Straughan}, \bibinfo{title}{Stability and Wave Motion in
  Porous Media}, \bibinfo{publisher}{Springer}, \bibinfo{address}{New York,
  NY}, \bibinfo{year}{2008}.
\bibitem[{Nield and Bejan(2017)}]{NiBe17}
\bibinfo{author}{D.~A. Nield}, \bibinfo{author}{A.~Bejan},
  \bibinfo{title}{Convection in Porous Media}, \bibinfo{edition}{5th} ed.,
  \bibinfo{publisher}{Springer}, \bibinfo{address}{New York, NY},
  \bibinfo{year}{2017}.
\bibitem[{Barletta(2022)}]{barletta2022boussinesq}
\bibinfo{author}{A.~Barletta},
\newblock \bibinfo{title}{The {B}oussinesq approximation for buoyant flows},
\newblock \bibinfo{journal}{Mechanics Research Communications}
  \bibinfo{volume}{124} (\bibinfo{year}{2022}) \bibinfo{pages}{103939}.
\bibitem[{Liang et~al.(2021)Liang, Tian, Wang, and Xu}]{liang2021non}
\bibinfo{author}{Y.~Liang}, \bibinfo{author}{P.~Tian},
  \bibinfo{author}{S.~Wang}, \bibinfo{author}{W.~Xu},
\newblock \bibinfo{title}{Non-{F}ickian diffusion in time-space fluctuating
  diffusivity landscapes: From superfast to ultraslow},
\newblock \bibinfo{journal}{Fractals} \bibinfo{volume}{29}
  (\bibinfo{year}{2021}) \bibinfo{pages}{2150191}.
\bibitem[{Girelli et~al.(2025)Girelli, Giantesio, Musesti, and
  Penta}]{girelli2025dynamical}
\bibinfo{author}{A.~Girelli}, \bibinfo{author}{G.~Giantesio},
  \bibinfo{author}{A.~Musesti}, \bibinfo{author}{R.~Penta},
\newblock \bibinfo{title}{Dynamical anomalous transport of molecules subject to
  inhomogeneous body forces},
\newblock \bibinfo{journal}{Zeitschrift f{\"u}r Angewandte Mathematik und
  Physik} \bibinfo{volume}{76} (\bibinfo{year}{2025}) \bibinfo{pages}{249}.
\bibitem[{Khalil(2002)}]{khalil2002nonlinear}
\bibinfo{author}{H.~K. Khalil}, \bibinfo{title}{Nonlinear Systems},
  \bibinfo{edition}{3rd} ed., \bibinfo{publisher}{Prentice Hall},
  \bibinfo{address}{Upper Saddle River, NJ}, \bibinfo{year}{2002}.

\end{thebibliography}

\end{document}